\documentclass[preprint,onecolumn,12pt]{IEEEtran}
\usepackage{blindtext}

\usepackage{mathtools}

\DeclarePairedDelimiter\floor{\lfloor}{\rfloor}
\usepackage[normalem]{ulem}

\usepackage{xcolor}
\usepackage{pdfsync}

\usepackage{graphicx}
\usepackage{epsfig}
\usepackage{amsmath}
\usepackage{amssymb}
\usepackage{amsthm}
\usepackage{mathtools}
\usepackage{amsfonts}
\usepackage{lineno}
\usepackage{tikz}

\usepackage{multirow}
\usepackage{arydshln}

\usetikzlibrary{positioning}
\usetikzlibrary{arrows}
\newtheorem{theorem}{Theorem}
\newtheorem{definition}[theorem]{Definition}

\newtheorem{assumption}[theorem]{Assumption}
\newtheorem{proposition}[theorem]{Proposition}
\newtheorem{remark}[theorem]{Remark}

\newcommand{\abs}[1]{\left|#1\right|}

\def\field#1{\mathbb #1}%
\def\R{\field{R}}%
\def\Z{\field{Z}}%
\newcommand{\Rn}[1][n]{\R^{#1}}

\newcommand{\Rp}{\R_{\geq 0}}
\newcommand{\Rsp}{\R_{> 0}}
\newcommand{\Zp}{\Z_{\geq 0}}
\newcommand{\Zsp}{\field{N}}

\def\K{\mathcal{K}}%

\def\KL{\mathcal{KL}}%
\def\Kinf{\mathcal{K}_\infty}%
\let\ol=\overline%
\let\ul=\underline%
\def\C{\mathcal{C}}%
\def\D{\mathcal{D}}%

\usepackage{enumerate}

\usepackage{hyperref}  

\def\nolabelhere{\leavevmode\ifmmode\nonumber\else\fi}%

\begin{document}

\title{Stability analysis of networked control systems with not necessarily UGES protocols}

\author{Seyed~Hossein~Mousavi,
	Navid~Noroozi, Anton~H.~J.~de~Ruiter, and~Roman~Geiselhart
	\thanks{S. H. Mousavi and A. H. J. de Ruiter are with the department of Aerospace Engineering, Ryrson University, Toronto, Ontario, Canada, emails: \texttt{\{seyedhossein.mousavi,aderuiter\}@ryerson.ca}.}
	\thanks{Navid Noroozi is with the %
		University of Passau, %
		Faculty of Computer Science and Mathematics, %
		Innstra\ss e 33, %
		94032 Passau, %
		Germany, email: \texttt{navid.noroozi@uni-passau.de}. The work of N. Noroozi was supported by the Alexander von Humboldt Foundation.}
	\thanks{Roman Geiselhart is with the University of Ulm, Institute of Measurement, Control and Microtechnology, Albert-Einstein-Allee 41, 89081 Ulm,
		Germany, email: \texttt{roman.geiselhart@uni-ulm.de}.}
}

\maketitle

\begin{abstract}
This note studies (practical) asymptotic stability of nonlinear networked control systems whose protocols are not necessarily uniformly globally exponentially stable.
In particular, we propose a Lyapunov-based approach to establish (practical) asymptotic stability
of the networked control systems.
Considering so-called modified Round Robin and Try-Once-Discard protocols, which are only uniformly globally
asymptotically stable, we explicitly construct Lyapunov functions for these two protocols, which fit our proposed setting.
In order to optimize the usage of communication resource, we exploit the following transmission
policy: wait for a certain minimum amount of time after the last sampling instant and then check a state-dependent criterion. When the latter condition is violated, a transmission
occurs. In that way, the existence of the minimum amount of time
between two consecutive transmission is established and so-called
Zeno phenomenon, therefore, is avoided.
Finally, illustrative examples are given to verify the effectiveness of our results.
\end{abstract}



\maketitle
\vspace{-1mm}
\section{Introduction}
Networked Control Systems (NCSs) are classes of systems where plant and controller are connected through a communication network. The importance to consider the underlying digital communication layer when designing and analyzing control systems has been widely recognized.

The so-called {\it emulation approach} is the most popular method in analysis and design of \emph{nonlinear} NCS~\cite{Walsh.2001,Nesic.2004,Carnevale.2007,Heemels.2010}.
In this approach, first, the communication network constraints are ignored and a continuous-time controller is designed for the continuous-time plant.
Then, it is shown that the stability and performance for network-based implementation of the system is maintained in an appropriate sense, if the data transmission frequency is sufficiently high {and the scheduling protocol meet some desired stability}. 
Emulation-based approaches are favorable because of their simplicity and also from the point that various standard tools in the continuous-time domain can still be applied through the controller design process.
Basically, knowing the maximum allowable transmission interval (MATI) is a key issue in the emulation design procedure of {\emph{time-triggered}} NCSs, since it quantifies the stability margin in terms of the transmission period.
As a seminal contribution, in~\cite{Nesic.2004} the authors develop a \emph{hybrid systems} framework for modeling and ${\cal L}_p$-stability analysis of nonlinear NCSs, where transmission and scheduling effects induced by a communication channel are taken into account.
In particular, an \emph{explicit} formula for computing an upper bound for the MATI is given in~\cite{Nesic.2004}.
In~\cite{Carnevale.2007}, the results in~\cite{Nesic.2004} have been improved by providing a \emph{less} conservative explicit upper bound of the MATI.
Over the last decade, the works~\cite{Nesic.2004,Carnevale.2007} have received great attention and further extensions toward handling other issues due to a digital network such as communication time-delays~\cite{Heemels.2010}, quantization effects~\cite{Nesic.2009b}, and reference tracking problems~\cite{Postoyan.2014}.
However, in all the mentioned works, only a class of scheduling protocols, including {\it classic} Round Robin (RR) and try-once-discard (TOD) protocols, is considered in which the Lyapunov function of the discrete-time system induced by the protocol has to \emph{exponentially} decay.

A central idea for optimizing the usage of communication resources is to lower data transmission over the communication network.
One way towards this end, is modifying existing standard protocols such that less often data is transmitted through the network. 
The authors in~\cite{Nesic.2004b} study stability of NCSs with protocols with a lower data transmission rate.
In particular, two {\it modified} versions of RR and TOD protocols are introduced in~\cite{Nesic.2004b}, which are shown \emph{not} to be uniformly globally exponentially stable (UGES) but uniformly globally asymptotically stable (UGAS).
However, no explicit formula for the computation of the MATI is provided in~\cite{Nesic.2004b}.

As the first contribution of this paper, we introduce Lyapunov-based conditions for stability analysis of NCSs, which not only relax the need for exponential decay-{rate} of the Lyapunov function associated with the discrete-time system induced by the protocol, but also provides an explicit upper bound for the MATI.
Note that the MATI formula proposed in~\cite{Carnevale.2007}, is developed based on the Lyapunov-based conditions that {demand an \emph{exponential} decay-rate of the Lyapunov function associated with the protocol; hence, they are particularly applicable to NCSs with UGES protocols.
It should be noted that, in the context of discrete-time systems,
any Lyapunov function associated with a globally asymptotically stable discrete-time system can also be scaled to satisfy the required exponential decay-rate assumption~\cite[Theorem 2.3]{Grune.2014}.
However, in general, it may not be an easy task to find such a scaling function. Also, in case the proper scaling function is found, the resulting Lyapunov function does not necessarily satisfy all the other conditions required in~\cite{Carnevale.2007}.}
{In this regard and} compared with~\cite{Carnevale.2007}, we only consider an \emph{asymptotic} decay-rate for the Lyapunov function associated with the protocol.
Consequently a wider class of protocols are incorporated, while an explicit formula for MATI is still provided.
{However, the reusling upper bound in this case could become very small.
In order to relax conservatism, the upper-bound for MATI is computed online at each transmission instant (see Theorem~\ref{thm:ncs-sgc} for more details).}

{As the second contribution of this work, we explicitly construct Lyapunov functions for the modified RR and TOD protocols. The resulting Lyapunov functions and the associated gain functions fit into the Lyapunov-based conditions given to establish the (practical) asymptotic stability of the NCSs.}
{We note that although the UGAS property of the modified TOD and RR are shown in~\cite{Nesic.2004b}, the results are not constructive, meaning that \emph{no} explicit Lyapunov function and the associated gains are provided.}

Another idea toward reduction of the rate of data transmission is to use {\it event-triggered} control mechanisms.
Generally speaking, in an event-based control system, data communication between plant and controller is scheduled using a pre-designed triggering condition.
In other words, in such systems data is not transmitted unless a specific triggering condition is violated.
It has been widely recognized an event-triggering mechanism is effectively able to optimize the usage of the network, while the desired stability and performance are preserved; see, \emph{e.g.},~\cite{Astrom.1999,Tabuada.2007}.

{Similar to~\cite{Dolk.2017,Abdelrahim.2016}, to further reduce the rate of data communication, we combine  an event-triggering condition called deadband control (\cite{Otanez.2002,Antsaklis.2013}) and the time-regularization subject to scheduling.
To be more precise, we exploit the following transmission policy: $i$) we first wait for $T$ units of times after the last sampling instant, where $T$ corresponds to the MATI already discussed earlier as the first contribution of this work; $ii$) we then check the deadband condition.
When the latter condition is violated, a transmission occurs.
In a deadband control mechanism, no new information is broadcast over the network, if the norm of the network-induced error ({\it i.e.} the difference between the last transmitted and current outputs values) lies within a certain deadband.
We finally note that from item $i$), the existence of the minimum amount of time between two consecutive transmission is established, which implies that so-called Zeno phenomenon~\cite{Goebel.2012} is avoided.}

The rest of this paper is organized as follows. Section~\ref{sec:Pre} and \ref{sec:problem-statement}, respectively, contain preliminaries and the problem statement. In Section~\ref{sec:protocols}, stability of the modified TOD and RR protocols is investigated.
Section~\ref{sec:main-results} proposes new conditions for stability analysis of NCSs with UGAS protocols as well as an event-triggering policy for further reduction of the data transmission rate.
Simulation results are given in Section~\ref{sec:example} and finally, concluding remarks are provided in Section~\ref{sec:conclusions}.

\vspace{-1mm}
\section{Preliminaries}\label{sec:Pre}

In this note, $\Rp (\Rsp)$ and $\Zp(\Zsp)$ are the nonnegative (positive) real and nonnegative (positive) integer numbers, respectively.
{For a set $\mathcal{S} \subset \Rn$, $\operatorname{cl}(\mathcal{S})$ denotes the closure of $\mathcal{S}$.}
The standard Euclidean norm is denoted by $\abs{\cdot}$.
{We denote the floor function by $\lfloor \cdot \rfloor$.}
We write $(x,y)$ to represent $[x^\top,y^\top]^\top$ for any pair $(x,y) \in \Rn\times \Rn[m]$.
The identity $n$ by $n$ matrix is denoted by $I_n$.
A function $\rho \colon \Rp \to \Rp$ is positive definite if it is continuous, zero at zero and positive elsewhere.
A positive definite function $\alpha$ is of class $\K$ ($\alpha \in \K$) if it is  strictly increasing.
It is of class $\Kinf$ ($\alpha \in \Kinf$) if $\alpha \in \K$ and also
$\alpha(s) \to \infty$ if $s \to \infty$.
A continuous function $\gamma \colon \Rp \to \Rp$ is of class $\mathcal{L}$ ($\gamma \in \mathcal{L}$) if it is decreasing and $\lim_{s \to \infty} \gamma (s) \to 0$.
A function $\beta \colon \Rp \times \Rp \to \Rp$ is of class $\KL$ ($\beta \in \KL$), if for each  $s \geq 0$, $\beta(\cdot,s) \in \K$, and for each $r \geq 0$, $\beta (r,\cdot) \in \mathcal{L}$.
A function $\beta \colon \Rp \times \Rp \times \Rp \to \Rp$ is of class $\mathcal{KLL}$ ($\beta \in \mathcal{KLL}$), if for each $s \geq 0$, $\beta (\cdot,s,\cdot) \in \KL$ and $\beta (\cdot,\cdot,s) \in \mathcal{KL}$.
For a locally Lipschitz function $U: \Rn \to \Rp$ and a vector $y \in \Rn$, $U^{\circ}(x;y):= {\limsup}_{h \to 0^+, z\to x}(U(z+hy)-U(z))/h$. For a continuously differentiable $U$, $U^{\circ}$ reduces to the standard directional derivative $\left\langle \nabla U(x),y \right\rangle$, where $\nabla U$ denotes gradient.

Due to lack of space, the basics of hybrid dynamical systems are not presented here. The reader is referred to~\cite{Goebel.2012} for detailed information. 

\vspace{-1mm}
\section{Problem Statement} \label{sec:problem-statement}

Consider the nonlinear plant model
	\vspace{-1mm}
\begin{align} \label{eq:e01}
\dot{x}_p= f_p (x_p,u), \ \ y = g_p (x_p),
\end{align}
where $x_p \in \Rn[n_p]$ is the plant state, $u \in \Rn[n_u]$ is the control input, and $y \in \Rn[n_y]$ is the plant output. 

It is assumed that the plant is output feedback stabilizable and we know a continuous-time controller which globally asymptotically stabilizes the origin of system \eqref{eq:e01} in the absence of a packet-based communication network.
We focus on dynamic controllers of the form
	\vspace{-1mm}
\begin{align} \label{eq:e06n}
\dot{x}_c=f_c (x_c,y), \ \ u = g_c (x_c),
\end{align}
where $x_c \in \R^{n_c}$ is the controller state.

We consider the scenario where the plant and controller are connected via a packet-based communication network that is composed of $\ell \in \Zsp$ nodes.
Each node corresponds to a collection of sensors and/or actuators of the plant and the controller.

The network imposes different constraints on the communication of both $u$ and $y$.
In this paper, we concentrate on the effect arisen due to sampling and scheduling.
Data transmissions only happen at some time instants $t_j, j \in \Zsp$, satisfying $0\leq t_0 < t_1 < t_2 < \!\dots$.
The structure of the network is in a way that at each transmission instant, a single node is granted access to the network. This scheduling is carried out by the transmission policy.

The overall system can be modelled as the following impulsive system \cite{Nesic.2004}
\begin{align} \label{eq:e09n}
&\begin{array}{rcll}
 \dot{x}_p &=& f_p (x_p,\hat{u}) & t \in [t_{j-1},t_j] \\
 y &=& g_p (x_p) \\
 \dot{x}_c &=& f_c (x_c,\hat{y}) & t \in [t_{j-1},t_j] \\
 u &=& g_c (x_c) \\
{\dot{\hat{y}}} &=& \hat{f}_p (x_c,\hat y, \hat u) & t \in [t_{j-1},t_j]\\
 {\dot{\hat{u}}} &=& \hat{f}_c (x_p,\hat y, \hat u) & t \in [t_{j-1},t_j] \\
 \hat{y} (t^+_j) &=& y (t_j) + h_y (j,e(t_j)) \\
 \hat{u} (t^+_j) &=& u (t_j) + h_u (j,e(t_j)),
\end{array}
\end{align}
where $\hat{y} \in \R^{n_y}$ and $\hat{u} \in \R^{n_u}$ are, respectively,  the estimate of measurements at the controller side and the currently available estimates of the true controller output at the plant side. 
These two variables are generated by the holding functions $\hat f_p$ and $\hat f_c$ between two successive transmission instants.
For the case that zero-order-hold devices are exploited, we would have $\hat f_p = 0$ and $\hat f_c = 0$.
The functions $h_y$ and $h_u$, respectively, accommodate the effect of the transmission protocol on the updates of $\hat y (t_j)$ and $\hat u (t_j)$ at the transmission time $t_j$.
Moreover, $e := (e_y,e_u) \in \R^{n_e}$ represents the network-induced errors in which $e_y := \hat{y} - y\in\Rn[n_y]$ and $e_u := \hat{u} - u\in\Rn[n_u]$.
Given $x := (x_p,x_c) \in \R^{n_x}, h := (h_y,h_u) \in \Rn[n_u + n_y]$, we rewrite \eqref{eq:e09n} as
	\vspace{-1mm}
\begin{subequations} \label{eq:e05}
\begin{eqnarray}
\dot{x} &=& f (x,e), \label{eq:e05a} \\
\dot{e} &=& g (x,e), \label{eq:e05b} \\
e (t^+_j) &=& h (j,e(t_j)), \label{eq:e05c}
\end{eqnarray}
\end{subequations}
where $f \colon \R^{n_x} \times \R^{n_e} \to \R^{n_x}$ and $g \colon \R^{n_x} \times \R^{n_e} \to \R^{n_e} $ are defined by $f (x,e) :=  \big( f_p (x_p , g_c (x_c) + e_u),  
f_c (x_c, g_p(x_p) + e_y)\big)$, $g (x,e) :=  \big( g_1(x,e),  g_2(x,e)\big)$
with $g_1(x,e) :=  \hat{f}_p (x_p,x_c,g_p (x_p) + e_y, g_c (x_c) + e_u) - \frac{\partial g_p}{\partial x_p} (x_p) f_p (x_p , g_c(x_p)+e_u)$ and
 $g_2(x,e) := \hat{f}_c (x_p,x_c,g_p(x_p) + e_y, g_c(x_c) + e_u) -\frac{\partial g_c}{\partial x_c} (x_c) f_c (x_c , g_p(x_p)+e_y).$ 
The function $h$ in~\eqref{eq:e05c} is called the scheduling protocol.

In order to study the protocol stability properties, the {\it protocol-induced} discrete-time system is defined by
	\vspace{-1mm}
\begin{equation}\label{eq:discrete_prot}
e(i+1) = h(i,e(i)).
\end{equation}

\begin{definition} \label{D:UGAS}
The discrete-time system \eqref{eq:discrete_prot} is uniformly globally asymptotically stable (UGAS) with a Lyapunov function $W \colon \Zp \times \Rn[n_e]  \to \Rp$ if there exist $\ul\alpha_e,\ol\alpha_e \in \Kinf$  and $\sigma \in \K$ with {$\sigma (s) < s$ for all $s \in (0,\infty)$} such that for all $j \in \Zp, e \in \R^{n_e}$ the following holds
	\vspace{-1mm}
	\begin{eqnarray}
	\ul\alpha_e (\abs{e}) & \leq & W(j,e) \leq \ol\alpha_e (\abs{e}), \label{eq:de18}  \\
	W(j+1,h(j,e)) & \leq & \sigma(W(j,e))  . \label{eq:de19}
	\end{eqnarray}
\end{definition}
In~\cite{Carnevale.2007,Nesic.2004}, Lyapunov-based conditions are imposed on protocol-induced system~\eqref{eq:discrete_prot}, {where the function $\sigma$ in~\eqref{eq:de19} needs to be linear.} Hence, the conditions are specifically applicable to uniformly globally exponentially stable (UGES) protocols.
As shown in~\cite{Nesic.2004b}, not every protocol is UGES.
The modified TOD and RR are two examples of such protocols, which are described in more details in Section \ref{sec:protocols}.

In this note, we propose a transmission policy which has two different phases depending on the value of the network-induced error $e$ and a predefined deadband $d \in \Rp$.
Define
	\vspace{-1mm}
\begin{equation}\label{eq:Ed}
{E_d := \{ e\in \R^{n_e}: |e_i|\leq \frac{d}{\ell}, \  \text{for} \ i=1,\cdots,\ell \}}.
\end{equation}
{The proposed transmission paradigm is given as follows: For $T_j > 0, j \in \Zsp$ units of times after the last transmission instant (\emph{i.e.} $t_j$) \emph{no} data is transmitted over the network, then}
\begin{itemize}
	\item if $e \in E_d$, no data is still transmitted over the network.
	Basically, this condition prevents data transmission when there is no remarkable change (quantified by $d$) in the nodes' output information.
	\item if $e \notin E_d$, this conditions holds, data transmission is scheduled by a UGAS protocol. 
\end{itemize}
{We note that Zeno phenomenon is avoided by the use of the proposed transmission paradigm.}
We aim to study the stability properties of the NCS under the above-explained network policy as well as provide an explicit formula for computing {$\{T_j\}_{j \in \Zp}$} such that the desired stability property is achieved.

\vspace{-1mm}	
\section{UGAS Protocols}\label{sec:protocols}
Here we examine the modified versions of TOD and RR originally introduced in~\cite{Nesic.2004b}.
Propositions~\ref{P:TOD} and \ref{P:RR} below \emph{explicitly} construct Lyapunov functions satisfying (\ref{eq:de18}) and (\ref{eq:de19}) for the modified TOD and RR protocols, respectively.
The proofs of Propositions~\ref{P:TOD} and \ref{P:RR} can be found in Appendix.

\vspace{-3mm}
\subsection{Modified TOD} 
This protocol behaves for large $e$ in the same way as TOD protocol, but for small $e$ the error jumps are smaller.
In this case the transmission protocol is defined by
	\vspace{-1mm}
\begin{align}\label{eq:tod}
h (i,e) = (I - \Psi(e) ) e ,
\end{align}
where $\Psi(e) := \mathrm{diag}(\psi_1(e) I_{n_1},\psi_2(e) I_{n_2},\dots,\psi_\ell(e) I_{n_\ell})$, and for each $j \in \{1,2,\dots,\ell \}$
\begin{align}\label{eq:tod-psi}
&\!\!\! \psi_j (e) := \left\{\begin{array}{lcl}
\!\mathrm{sat} (\abs{e_j}) & \! \mathrm{if}\! & \!j\! =\! \min (\mathrm{arg} \max_j \abs{e_j}), \\
\!0 & & \!\!\!\!\!\!\!\!\mathrm{otherwise}, 
\end{array}\right.
\end{align}
with $\mathrm{sat} (s) := \min \{s,1\}$ for all $s \geq 0$.
As discussed in \cite{Nesic.2004b}, this protocol is \emph{not} UGES and transmits less data compared to its original version.
\begin{remark}
The original TOD protocol is also described by \eqref{eq:tod}, but $\psi_j (e)$ is defined as  
  \begin{align}\label{eq:tod-psi-trad}
  \psi_j (e) := \left\{\begin{array}{lcl}
  1 & \mathrm{if} & j = \min (\mathrm{arg} \max_j \abs{e_j}), \\
  0 & & \mathrm{otherwise}. 
  \end{array}\right.
  \end{align}	
\end{remark}
\begin{proposition}\label{P:TOD}
The modified TOD protocol is UGAS with the Lyapunov function $|e|$ with the associated gain functions $\ul \alpha_e(s) = \ol \alpha_e(s) = s$, and 
\begin{align} \label{eq:sigmaTOD}
\sigma(s) =  \left\{ \begin{array}{lcl}
s \sqrt{1 - {\ell^{-3/2}} s } & \mathrm{if} & 0\leq s \leq \sqrt{\ell}, \\
\sqrt{({\ell -1})/{\ell}} s & \mathrm{if} & s > \sqrt{\ell}.
\end{array} \right. 
\end{align}	
\end{proposition}
\vspace{-3mm}
\subsection{Modified RR}
This protocol behaves exactly in the same way as RR for large error $e$, but it transmits less frequently for small $e$. In this case the transmission policy is defined as
\begin{equation}\label{MRR}
h(i,e) = (I - \Delta(i,e))e,
\end{equation}
where $\Delta(i,e) = \text{diag} \{ \delta_1(i,e) I_{n_1},\cdots,\delta_{\ell} I_{n_{\ell}} \}$, and 
\begin{equation}\label{RR-delta}
\delta_k(i,e)\! :=\!\left\{\begin{array}{lcl}
1 & \mathrm{if} &\! \!\!\abs{e}>0, i = \floor{\frac{1}{\mathrm{sat} (\abs{e})}}(k+j \ell),j \in \mathbb{N},  \\
0 & & \!\!\!\mathrm{otherwise}. 
\end{array}\right.
\end{equation}
{Basically, $\delta_i$ switches on and off the action corresponding to node $i$ in the policy equation \eqref{MRR}}. 
	\vspace{-1mm}
\begin{remark}
	The original RR protocol is also described by~\eqref{MRR}, but $\delta_k(i,e)$ is defined as  
\begin{equation}\label{RR-delta-trad}
\delta_k(i,e) :=\left\{\begin{array}{lcl}
1 & \mathrm{if} & i = k+j \ell, j \in \mathbb{N},  \\
0 & & \mathrm{otherwise}. 
\end{array}\right.
\end{equation}
\end{remark}
\begin{proposition}\label{P:RR}
	The modified RR protocol is UGAS with the Lyapunov function $W(i,e) = \sqrt{\sum_{k=i}^{+\infty}{|\phi(k,i,e)|^2}}$, where $\phi(k,i,e)$ is a solution to \eqref{eq:discrete_prot} with $h$ given by \eqref{MRR}, at sample $k$, starting at initial time $i$ with initial condition $e$. Moreover the associated gain functions in \eqref{eq:de18} and \eqref{eq:de19} are $\ul\alpha_e(s)=s$,  $\ol\alpha_e(s)=\max \left\{ \ell \sqrt{2s}, \sqrt{\ell} s \right\}$, and
		\vspace{-1mm}
\begin{align} \label{eq:sigmaRR}
\sigma(s) =  \left\{ \begin{array}{lcl}
s \sqrt{1 - {s^2}/({4\ell^4}}) & \mathrm{if} & 0\leq s < 2 \sqrt{\ell^3}, \\
\sqrt{{(\ell -1)}/{\ell}} s & \mathrm{if} & s \geq 2 \sqrt{\ell^3}.
\end{array} \right.
\end{align}
\end{proposition} 
We note that \eqref{eq:sigmaTOD} and \eqref{eq:sigmaRR} are nonlinear functions, which are not upper bounded by a linear function less than the identity function.
In particular, the modified TOD and RR are \emph{not} UGES but proved to be UGAS.
Although these two protocols are studied in~\cite{Nesic.2004b}, no \emph{explicit} Lyapunov and associated gain functions  are provided therein. 
  
\section{Stability Properties of the NCS with UGAS Protocols} \label{sec:main-results}
In this paper, we provide an explicit state-dependent bound for MATI, which guarantees global (practical) asymptotic stability of the proposed NCS.
Toward this end, we first transform the NCS model~\eqref{eq:e05} into a hybrid system such that the analytical tools of \cite{Goebel.2012} can be exploited to infer the stability properties of the system.
In particular, we introduce the auxiliary clock variable $\tau \in \Rp$ representing the time elapsed since the last transmission instant.
We also introduce $\kappa \in \Zp$ to count the number of transmissions. Denote $\xi := (x,e,\tau,\kappa)$, $F(\xi) :=(f(x,e),g(x,e),1,0)$ and $G(\xi) := (x,h(\kappa,e),0,\kappa+1)$.
The following hybrid system representation of the NCS~\eqref{eq:e05} is obtained
	\vspace{-1mm}
\begin{align} \label{ncs-hybrid}
& \mathcal{H}_\mathrm{NCS} := \left\{ \begin{array}{l}
 \dot{\xi} = F (\xi) \;\; \quad \xi \in \mathcal{C}, \\
  \xi^+ = G (\xi) \quad \xi \in \mathcal{D},
 \end{array} \right. &
\end{align}
\vspace{-1mm}
$\!\!$where $\C$ and $\D$ denote the flow and jump sets respectively. According to the network data transmission paradigm described in Section~\ref{sec:problem-statement}, these sets are defined by $\C$ $=$ $\{ (x,e,\tau,\kappa) \colon$ $\tau \in [0,T_j] \ \mathrm{or} \ {e \in E_d} \}$ and $\D = \{ (x,e,\tau,\kappa) \colon$ $ { \tau \geq T_j} \ \mathrm{and} \ {e \in \tilde E_d} \}$, {where $\tilde E_d := \operatorname{cl} \big(\R^{n_e} \backslash E_d\big)$}.

In the sequel, a Lyapunov-based approach for (practical) asymptotic stability of the NCS is presented.
To this end, { and similar to \cite{Carnevale.2007} and \cite{Postoyan.2014}}, we make the following assumptions.

\begin{assumption} \label{A:01}
	Consider the NCS~\eqref{ncs-hybrid}. There exist a locally Lipschitz function $V \colon \R^{n_x} \to \Rp$, a function $W \colon \Zp \times \R^{n_e} \to \Rp$ {that is locally Lipschitz in its second argument}, a continuous function $H \colon \R^{n_x} \to \Rp$, $\ul\alpha_x,\ol\alpha_x,\ul\alpha_e,\ol\alpha_e \in \Kinf, \sigma \in \K$ with $\sigma(s) < s$ for $s > \ul\alpha_e(d)$, and real numbers $L,\gamma > 0$, and $\eta >0$ such that for all $x \in \R^{n_x}$ it holds
		\vspace{-1mm}
	\begin{align}
	& \underline\alpha_x (\abs{x}) \leq V (x) \leq \overline\alpha_x (\abs{x}), & \label{eq:e17} 
	\end{align}
and for almost all $x \in \R^{n_x}$, for all $e \in \R^{n_e}$ and all $\kappa
	\in \Zp$ it holds that
	\vspace{-1mm}
	\begin{align}
		\langle \nabla V (x) , f(x,e) \rangle \leq & - \eta V(x) - \eta W^2(\kappa,e) - H^2(x)  + \gamma^2 W^2 (\kappa,e) . \label{eq:e18}
	\end{align}
	Moreover, we have
	\begin{flalign}
	& \!\!\!\!\!\ul\alpha_e (\abs{e})  \leq W (\kappa,e) \leq \ol\alpha_e (\abs{e}),\,\,  \forall \kappa \in \Zp , \forall e \in \R^{n_e},  \label{eq:e19} \\	
	& \!\!\!\!\!W (\kappa+1,h(\kappa,e))\! \leq\! \sigma \big( W (\kappa,e) \big), \ \ \forall \kappa \in \Zp , \forall e \in \tilde E_d; \label{eq:e20}
	\end{flalign}
for almost all $e \in \R^{n_e}$, all $x \in \R^{n_x}$ and all $\kappa \in \Zp$	
		\vspace{-1mm}
	\begin{equation}
	\left\langle  {\partial W (\kappa,e)}/{\partial e} , g(x,e) \right\rangle \leq \!L W (\kappa,e) \!+\! H(x) .  \label{eq:e21}
	\end{equation}
\end{assumption}
From equations~\eqref{eq:e17} and~\eqref{eq:e18}, the emulated controller assures an ISS-like property for the system $\dot x = f(x,e)$ with $W(\kappa,e)$ as input.
Different classes of linear and nonlinear systems satisfy~\eqref{eq:e17} and~\eqref{eq:e18} (see~\cite{Postoyan.2014}).
{From conditions~\eqref{eq:e19} and~(\ref{eq:e20}) the protocol only needs to be uniformly globally \emph{practically} stable.}

Let the parameters $\gamma, L$ come from Assumption~\ref{A:01}, {$\lambda \in (0,1)$} and define the notation
\begin{equation} \label{eq:MATIbound}
\small
\mathcal{T}(\gamma,L,\lambda) := \!\!
\left\{ \begin{array}{ll}
\!\!\frac{1}{L r}\tan^{-1}\left(\frac{r(1-\lambda)}{2\frac{\lambda}{\lambda+1} (\frac{\gamma}{L}-1) +1+\lambda}\right) & \gamma>L, \\
\!\!\frac{1}{L} \left( \frac{1-\lambda}{1+\lambda} \right) & L = \gamma, \\
\!\!\frac{1}{L r}\tanh^{-1}\left(\frac{r(1-\lambda)}{2\frac{\lambda}{\lambda+1} (\frac{\gamma}{L}-1) +1+\lambda}\right) & \gamma<L,
\end{array}\right. 
\end{equation}
where $r := \sqrt{\abs{(\gamma / L)^2-1}}$.
Theorem \ref{thm:ncs-sgc} contains the main result of this note. The proof is provided in Appendix.

\begin{theorem} \label{thm:ncs-sgc}
Consider system~(\ref{ncs-hybrid}) and let Assumptions~\ref{A:01} hold.
Generate a sequence $\{\lambda_{j}\}_{j \in \Zp}$ with $\lambda_{j} \in (0,1)$ for all $j \in \Zp$ as follows: $\lambda_{0} \geq \sigma(W(0,e_0))/{W(0,e_0)}$, $e_0 := e(0,0)$ and at all the other transmission instants $(t_{j},j-1) \in E$ with $E$ as the hybrid time domain, $\lambda_{j} \geq \sigma(W(\kappa(t_{j},j-1),e(t_{j},j-1)))/W(\kappa(t_{j},j-1),e(t_{j},j-1))$.
Accordingly, generate $\{T_{j}\}_{j \in \Zp}$, where {$T_j \in (0,\mathcal{T}(\gamma,L,\lambda_j)]$}.
There exist $\beta \in \mathcal{KLL}$ and $\delta > 0$ such that for all $x_0 := x(0,0) \in \R^{n_x}$, all $e_0 \in \Rn[n_e]$, all $\kappa (0,0) \in \Zp$, all $\tau (0,0) \in \Rp$ and  all $(t,j) \in E$ a solution of~\eqref{ncs-hybrid} satisfies
		\vspace{-1mm}
\begin{equation} \label{eq:GApS}
\abs{x(t,j)} \leq   \max \big\{ \beta \left(\abs{(x_0,e_0)},t,  j \right) , \delta \big\}, 
\end{equation}
with $\delta \leq \frac{ 2\lambda_{\max} (\gamma^2 - \eta)\overline \alpha_e^2(d)}{\lambda_{\min} \eta (1-e^{-\eta \epsilon})}$.
\end{theorem}
{From condition~(\ref{eq:e20}), we can always generate a sequence $\{\lambda_{j}\}_{j \in \Zp}$ satisfying the respective statement of Theorem~\ref{thm:ncs-sgc}.}

Theorem~\ref{thm:ncs-sgc} provides an explicit upper bound for {$T_j$}, assuring practical \emph{asymptotic} stability property in the sense of~\eqref{eq:GApS}.
Moreover, an explicit relation between the \emph{quality-of-control} (QoC) parameter $\delta$ (which quantifies the {the ultimate bound of states convergence}) and the \emph{quality-of-service} (QoS) deadband parameter $d$ (affecting the data transmission rate) is provided.
Hence, there is an explicit tradeoff between the system performance of the channel resources usage.
In particular, the deadband parameter $d$ can be used for resource-aware control design in NCSs. 
For more details on tradeoffs between QoC and QoS, see~\cite{Borgers2017}.

Although in~\cite{Nesic.2004b} the stability property of {\emph{time-triggered}} NCSs with not necessarily UGES protocols is studied, {\it no} explicit formula for the computation of MATI is provided.
On the contrary, Theorem~\ref{thm:ncs-sgc} proposes {an explicit upper bound for the minimum time between two consecutive transmission instants (\emph{i.e.} $T_j$), which can be viewed as MATI when only time-regularization subject to scheduling is exploited (\emph{i.e.} the deadband event-triggering mechanism is bypassed).}

{We note that $T_j$ is computed at each transmission instant and it determines the minimum amount of  time elapsed until the next transmission occurs.
More specifically, at each transmission instant $t_{j}$, $\mathcal{T}(\gamma,L,\lambda_{j})$ is calculated, where $\lambda_{j}$ is generated by Theorem~\ref{thm:ncs-sgc}.
Then we take $T_j \in (0, \mathcal{T}(\gamma,L,\lambda_{j})]$.
Finally the next transmission instant $t_{j+1}$ is scheduled according to the transmission policy: after $T_j$ unit of times the triggering condition { $\abs{e_i}\leq d/\ell$} for all $i \in \{1,\dots,\ell \}$ is continuously evaluated until it is violated.
}
\begin{remark}\label{rrrr}
Theorem~\ref{thm:ncs-sgc} is also valid if $T_j$ for all $j\in\Zp$ is computed for a fixed value of $\lambda_\mathrm{max}$, where $\lambda_\mathrm{max}$ is defined as $\lambda_\mathrm{max} := \max\limits_j \lambda_{j}$.
However, the state-dependent strategy provides less conservative values for $T_j$, compared to the fixed one.
See Section~\ref{sec:example} below for an illustrative example.
\end{remark}

\begin{remark}
{Basically, at each transmission instant, $T_j$ is updated based on the Lyapunov function decay rate ($\lambda_{j}^{-1}$)}.
From~\eqref{eq:sigmaTOD} the value computed for $\lambda_{j}$ (generated by Theorem~\ref{thm:ncs-sgc}) in the \emph{modified} TOD for $W(\kappa,e) > \sqrt{\ell}$ is $\sqrt{\frac{\ell-1}{\ell}}$, which is the same as the one for the \emph{classic} TOD (see~\cite{Nesic.2004}).
{This results in the same upper bound for $T_j$ for both protocols in this region}.
However, for $W(\kappa,e) \leq \sqrt{\ell}$, as $W(\kappa,e)$ decreases, $\lambda_{j}$ approaches 1. {Consequently, the Lyapunov function decay rate decreases, which in the view of~\eqref{eq:MATIbound} leads to a tighter upper bound for $T_j$ for the modified TOD.}
Similar arguments hold for the modified RR.
\end{remark}

\section{Illustrative Example} \label{sec:example}
The proposed approach is verified by applying to a batch reactor system which is a benchmark example in the NCS literature (see~\cite{Nesic.2004,Carnevale.2007,Heemels.2010}).
The reactor dynamics are modeled by the linear equations $\dot x_p = A_p x + B_p u$ and $y = C_p x$, in which $n_p=4$, $n_u=2$, $n_y=2$.
The dynamic controller is of the form $\dot x_c = A_c x_c + B_c y$ and  $u=C_c x_c$ with $n_c=2$.
We assume that only the plant outputs are sent to the controller over the network and so $\ell =2$. Using these parameters, functions $f$ and $g$ in \eqref{eq:e05} are formulated as $f(x,e) = A_{11} x + A_{12} e$ and $g(x,e) = A_{21} x + A_{22} e$, where the values for $A_{11}, A_{12}, A_{21}$ and $A_{22}$ are given in \cite{Heemels.2010}. Two different transmission protocols are considered and examined for this setup.
\subsection{Modified TOD}
\begin{figure}[ht!]
	\centering
	\includegraphics[width=.6\textwidth , height=5cm]{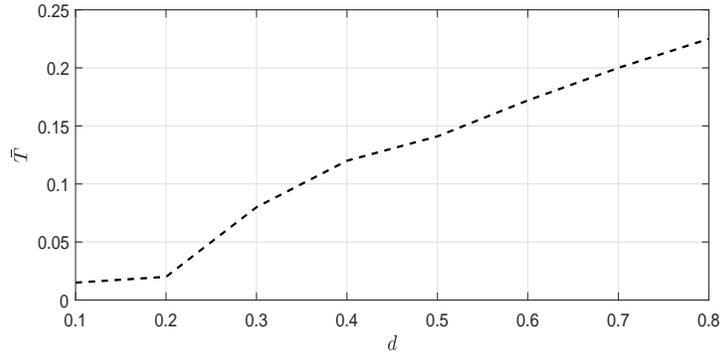}
	\vspace{-1mm}
	\caption{Average of transmission intervals as a function of the deadband $d$, while using modified TOD protocol, {with varying $\lambda_j$}.}
	\label{fig:tau_vs_d}
\vspace{-1mm}
\end{figure}

\begin{figure}
\centering
	\includegraphics[height=8cm,width=.8\textwidth]{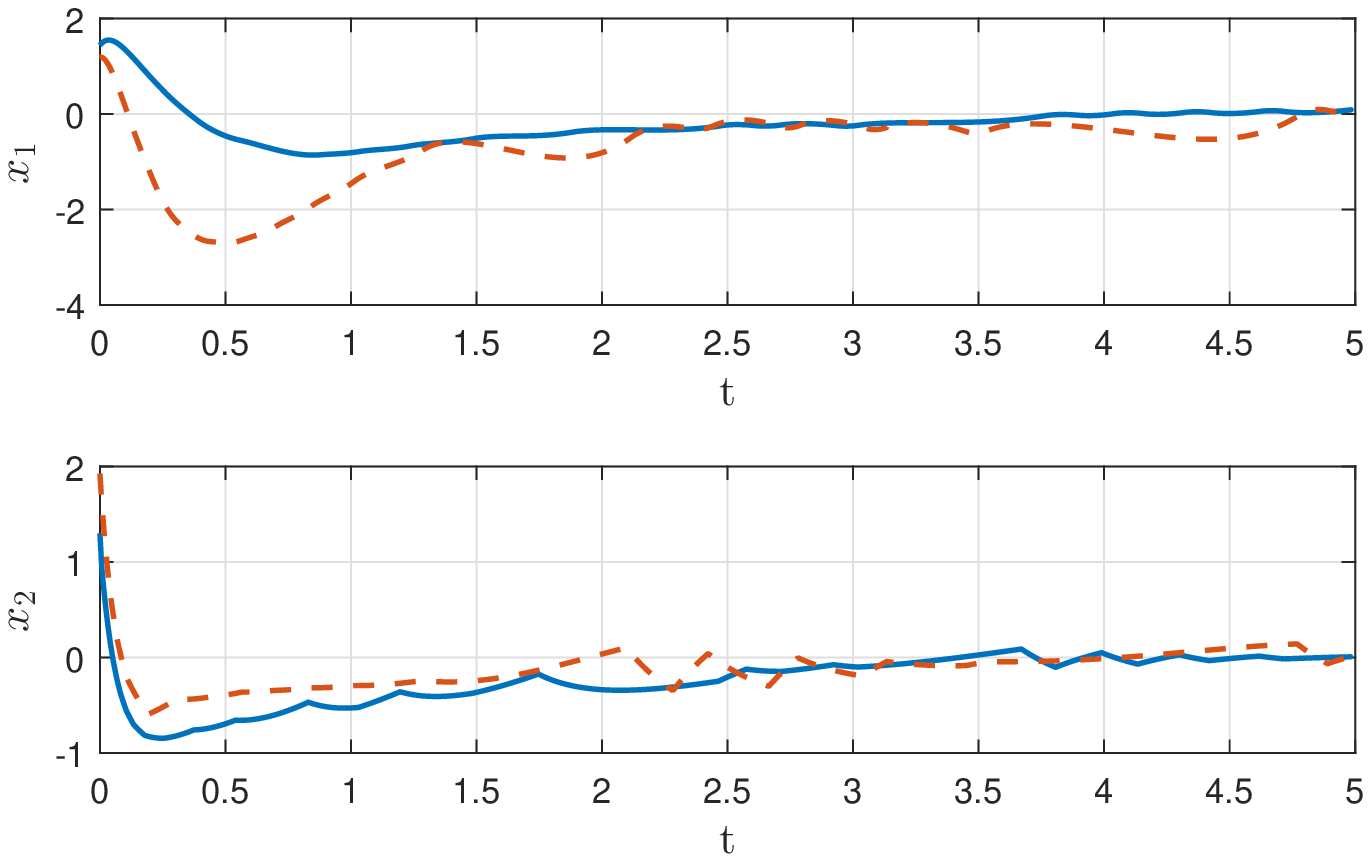}
\vspace{-1mm}
	\caption{Trajectories for the first and second states of the plant, for the cases $d=0.1$ (solid) and $d=0.6$ (dashed), while using modified TOD protocol.}
	\vspace{-1mm}
		\label{fig:statesTOD}
\end{figure}

\begin{figure}[ht!]
	\centering
	\includegraphics[width=.6\textwidth , height=5cm]{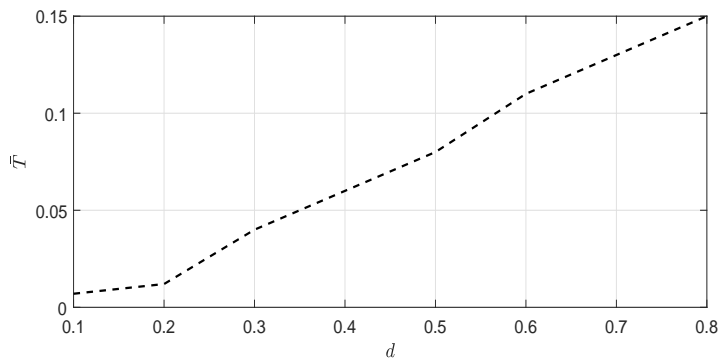}
	\vspace{-2mm}
	\caption{Average of transmission intervals as a function of the deadband $d$ , while using the modified TOD protocol, {with a fixed $\lambda_j = \lambda_\mathrm{max}$} for all $j\in \Zp$.}
	\label{fig:tau_vs_d_cons}      
	\vspace{-1mm}
\end{figure}

In this case, the transmission policy $h$ in \eqref{eq:e05} is described by \eqref{eq:tod} and \eqref{eq:tod-psi}.
The data transmission deadband $d$ is varied in the interval $[0.1,0.8]$ to illustrate its effectiveness on data communication reduction.
To verify Assumption \ref{A:01}, define $V(x) = x^T P x$, where $P$ is a positive definite matrix, and  $W(i,e)=|e|$.
Taking $H(x) = M|A_{12}x|$ and $L = M|A_{22}|$ (with $|\frac{\partial W}{\partial e}| \leq M = 1$),~\eqref{eq:e21} is clearly satisfied and $L$ is obtained as $L=15.73$.
Computing the derivative of $V$, one can show that~\eqref{eq:e18} holds if one can solve the following linear matrix inequalities (LMIs) for some $\epsilon > 0$
\begin{equation}\label{eq:LMIs}
\begin{array}{l}
\small
\! \! \! \! \! \! \!  \begin{bmatrix}
A_{11}^T P + P A_{11} + M^2 A_{21}^T A_{21} + \epsilon I & P A_{12} \\
A_{12}^T P    & (\epsilon-\gamma^2) I
\end{bmatrix} \preceq 0 .
\end{array}
\end{equation}
{From~\eqref{eq:MATIbound}, the smaller $\gamma$ is, the larger $T_j$ is obtained.}
Let $\epsilon = 0.001$ and solve LMIs~\eqref{eq:LMIs} with the objective of minimizing $\gamma^2$.
This gives $\gamma =16.92$.
For the simulation, $T_j$ is updated as $T_j =\Gamma(\gamma,L,\lambda_{j})$, where $\lambda_{j}$ is generated by Theorem~\ref{thm:ncs-sgc} and~\eqref{eq:sigmaTOD}.
Simulations are carried out for different values of $d \in [0.1,0.8]$ and initial conditions (randomly) satisfying $\abs{x_0}<2$ and $\abs{e_0}<1.5$.
Fig.~\ref{fig:tau_vs_d} shows that the average values of transmission intervals, denoted by $\ol T$, versus the deadband $d \in [0.1, 0.8]$ follows a trend, ascending from $20$ ms to $230$ ms.
However, as expected, this happens at the expense of a {larger ultimate bound of states convergence}.
This is shown by Fig.~\ref{fig:statesTOD}.
From Remark~\ref{rrrr}, we may use a fixed $T_j$ for all $j\in\Zp$, i.e. $\underline T := \mathcal{T}(\gamma,L,\lambda_\mathrm{max})$.
Table~\ref{tab:MATIs} reports $\underline T$'s for different values of $d$ varying from $0.1$ to $0.8$.
In this case, values for $\overline{T}$ are also obtained, as depicted by Fig.~\ref{fig:tau_vs_d_cons}.
Obviously, smaller $\overline{T}$ in this case are considerably smaller than those in~Fig.\ref{fig:tau_vs_d}.


\begin{table}[ht!]
\caption{Bounds for $\underline T$ for different values of $d$}
			\label{tab:MATIs}
	\centering
	\resizebox{13cm}{!}{
		\begin{tabular}{lcccccccc}
			\hline
			& \multicolumn{8}{c}{$d$} \\ \cdashline{2-9} 
			&  0.1 & 0.2 & 0.3 & 0.4 & 0.5 & 0.6 & 0.7 & 0.8 \\ \cline{1-9} 
			$\underline T$  & 0.0005 & 0.0011 & 0.0017 & 0.0023 & 0.0030 & 0.0036 & 0.0043 & 0.0051  \\
			\hline 
			\end{tabular}}
\end{table}

\subsection{Modified RR}
\begin{figure}
	\centering
	\includegraphics[width=.6\textwidth , height=5cm]{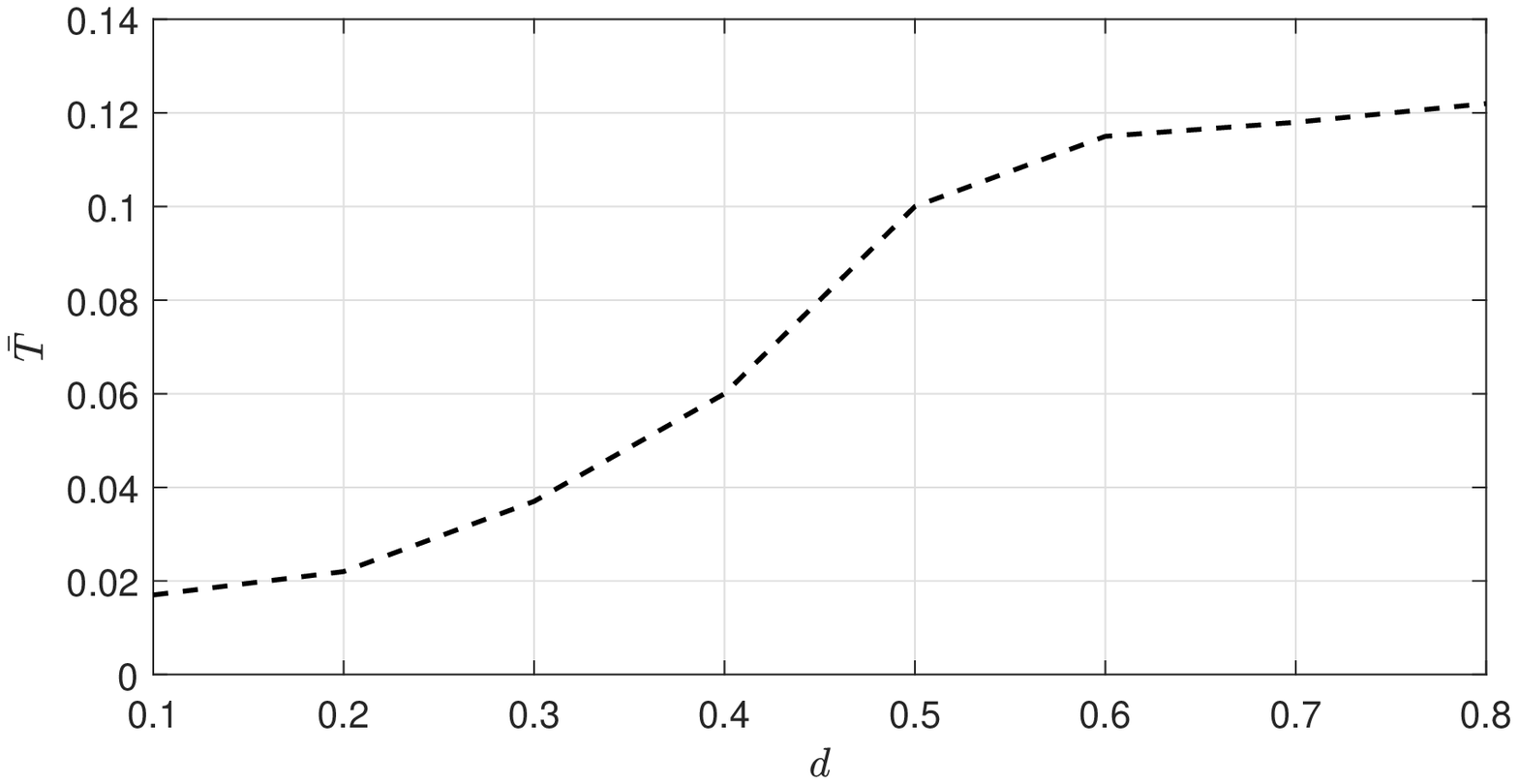}
	\caption{ Average of transmission intervals as a function of the deadband $d$, while using modified RR protocol.}      \label{fig:tau_vs_d_RR}
\end{figure}

\begin{figure}
	\centering
	\includegraphics[width=.8\textwidth , height=8cm]{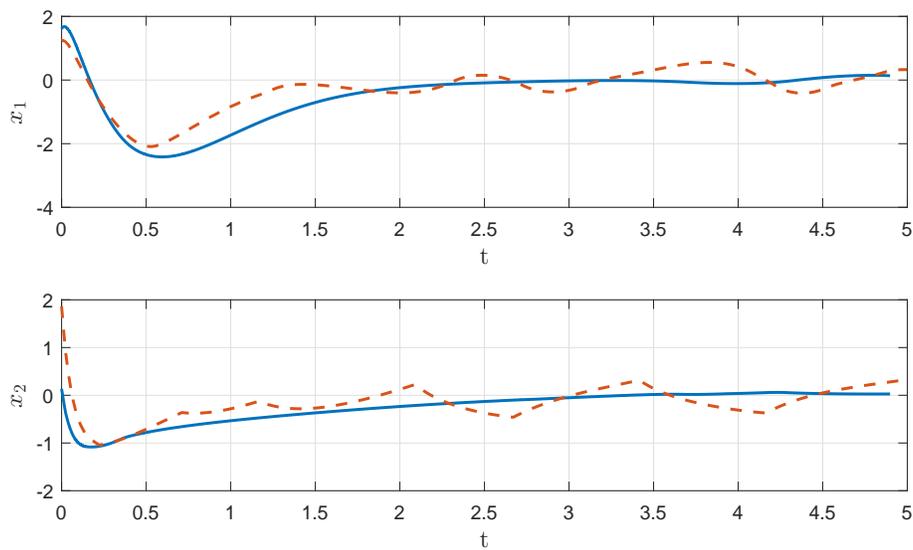}
	\caption{ Trajectories for the first and second states of the plant, for the cases $d=0.1$ (solid) and $d=0.6$ (dashed), while using modified RR protocol.}      \label{fig:statesRR}
\end{figure}

In this case the transmission protocol $h$ is modeled by \eqref{MRR} and \eqref{RR-delta}. Using the formulation for $L$ given in the previous part and setting $M$ to $\sqrt{\ell}$ (which is obtained using a similar approach as \cite{Nesic.2004}), $L$ is calculated as $L=22.24$.
Having solved the LMIs~\eqref{eq:LMIs} with the objective of minimizing $\gamma$, we obtain $\gamma=23.93$.
The rest of parameters are set similar to the previous part and simulations are done for different values of $d$. The results for the average values of $T_j$ versus $d$ are depicted in Fig~\ref{fig:tau_vs_d_RR}.
Moreover, the trajectories for the first and second states of the plant are shown for two different deadband values in Fig.~\ref{fig:statesRR}.

\section{Conclusions}\label{sec:conclusions}

The stability of a class of NCSs with not necessarily UGES protocols was investigated.
A set of UGAS protocols Lyapunov conditions was provided, and particularly, modified versions of RR and TOD protocols were proved to satisfy the conditions.
Moreover, an explicit formula for computation of the upper bound for the minimum amount of time between two consecutive transmission was provided.
Simulation results were provided to show the effectiveness of the approach.


\appendix

{\bf Proof of Proposition~\ref{P:TOD}}. Consider arbitrary $e$ and suppose without loss of generality that $\abs{e_1} = \max_j \abs{e_j}$. Then we would have $\abs{e_1}^2 \geq \frac{1}{\ell} \abs{e}^2$. Now consider two cases:
	i) If $\abs{e_1} \geq 1$, we have with the same arguments as those in \cite[Section IV]{Nesic.2004} that $W (h(e)) \leq \sqrt{{(\ell -1)}/{\ell}} \  W(e)$.
	ii) Assume that $\abs{e_1} < 1$. It follows from~\eqref{eq:tod-psi} that $\abs{h(e)}^2 = \sum_{j=2}^\ell \abs{e_j}^2 + (1-\abs{e_1})^2 \abs{e_1}^2 = \abs{e}^2 + \abs{e_1}^3 (\abs{e_1}-2)$.
Using the last equality and the facts $\abs{e_1} < 1$ and $\abs{e_1}^2 \geq \frac{1}{\ell} \abs{e}^2$, we have $W (h(e))  \leq \sqrt{\abs{e}^2 - \abs{e_1}^3 } \leq W (e) \sqrt{1 - \frac{1}{\ell^{3/2}} W(e)}$.

It follows from the results of cases i) and ii) that $W (h(e)) \leq \sigma (W(e))$,
where $\sigma : [0,+\infty) \to \Rp$ is defined by \eqref{eq:sigmaTOD}. Moreover, it is not hard to see that $\sigma \in \Kinf$ and therefore condition~\eqref{eq:de19} is satisfied. Trivially, \eqref{eq:de18} is also satisfied with $\ul\alpha_e(s)= \ol\alpha_e(s)= s$.
\hfill $\Box$

{\bf Proof of Proposition \ref{P:RR}}. 
Without loss of generality it is assumed that $|e_j|>0$ for all $j \in \{1,2,\dots,\ell \}$. Define vectors $e^{(i)} \in \Rn$ as follows: $e^{(i)}=0$ if $j \leq i$, and $e^{(i)}=0$ otherwise,
where $i \in \{1,2,\dots,\ell \}$ and $e^{(0)} := e$.
{Denote $m$ ($0 \leq m \leq \ell$) as the number of transmissions after which  the value of $|\phi(k,i,e)|$ drops below 1/2.} In the other words $|\phi(k,i,e)| \geq 1/2$ if $i \leq k \leq i+m-1$, and $|\phi(k,i,e)| < 1/2$ otherwise. {(Note that this always happens since if $|e_j|>1/2$ for all $j=1,\dots, \ell$, then, based on the definition of the protocol, the modified RR turns into the classic RR and the error vector becomes zero after $\ell$ steps; see \cite{Nesic.2004}.)}
Using this fact and based on the protocol definition given in (\ref{MRR}) and (\ref{RR-delta}), it is derived that $\sum_{k=i}^{i+m}{|\phi(k,i,e)|^2} \! \leq \! m |e|^2$, and  $\sum_{k=i+m+1}^{+\infty}{|\phi(k,i,e)|^2} \! \leq \!  (\ell + m+1) \floor{\frac{1}{\abs{e^{(m+1)}}}}|e^{(m+1)}|^2 + \dots  + (\ell + \ell - 1)\floor{\frac{1}{\abs{e^{(\ell-1)}}}}|e^{(\ell-1)}|^2 \leq (\ell - m) (\ell + \frac{\ell+m}{2})|e| \leq (\ell - m) (2\ell)|e|$.
Using these two inequalities we have $W(i,e) = \sqrt{\sum_{k=i}^{+\infty}{|\phi(k,i,e)|^2}}  \leq \sqrt{\frac{m}{\ell}(\ell |e|^2) + \frac{\ell-m}{\ell}(2 \ell^2|e|)}. $
As $\frac{m}{\ell} + \frac{\ell-m}{\ell} =1$ and $0 \leq m \leq \ell$, we get $W(i,e) \leq \max_{0\leq m \leq \ell}  \sqrt{\frac{m}{\ell}(\ell |e|^2) + \frac{\ell-m}{\ell}(2 \ell^2|e|)}$.
To find the max term in the last inequality, we consider two different cases: 
 i) If $\ell |e|^2 \leq 2 \ell^2 |e|$, the maximum happens at $m=0$, and so $W(i,e) \leq  \ell \sqrt{2|e|}$.
On the other hand, since $W(i,e) = \sqrt{\sum_{k=i}^{+\infty}{|\phi(k,i,e)|^2}}$, we have $W(i+1,h(i,e)) = \sqrt{W^2(i,e) - |e|^2}$.
It is obtained that $W(i+1,h(i,e)) \leq W(i,e)\sqrt{1 - \frac{W^2(i,e)}{4 \ell^4}}$. 
ii) If $\ell |e|^2 \geq 2 \ell^2 |e|$, the maximum happens at $m=\ell$, and so $W(i,e) \leq  \sqrt{\ell}|e|$. Using this inequality alongside with the fact $W(i+1,h(i,e)) = \sqrt{W^2(i,e) - |e|^2}$ results in $W(i+1,h(i,e)) \leq \sqrt{\frac{\ell-1}{\ell}} W(i,e)$.

It follows from the results of cases i) and ii) that $W(i+1,h(i,e)) \leq \sigma(W(i,e))$,
where $\sigma : [0,+\infty) \to \Rp$ is defined by \eqref{eq:sigmaRR}.
We note that $\sigma \in \Kinf$. Thus condition \eqref{eq:de19} is satisfied.
We also have $W(i,e) \leq \max \left\{ \ell \sqrt{2\abs{e}}, \sqrt{\ell} \abs{e} \right\}$.
On other hand, from the definition of $W(i,e)$ we have that $W(i,e) \geq \sqrt{|\phi(i,i,e)|^2} = \abs{e}$.
Thus condition \eqref{eq:de18} is satisfied with the functions $\ul\alpha_e(s)=s$ and $\ol\alpha_e(s)=\max \left\{ \ell \sqrt{2s}, \sqrt{\ell} s \right\}$. 
$\,$\hfill $\Box$

{\bf Proof of Theorem. \ref{thm:ncs-sgc}}
Let $T_j \leq \mathcal{T}(\gamma,L,\lambda_{j})$ be given.
Also, let the triple $(\gamma,L,\lambda_{j})$ generate $\phi_{j} \colon [0,T_j] \to \R$ via
\begin{align} \label{eq:e25}
\dot{\phi}_{j} = -2 L \phi_{j} - \gamma (\phi_{j}^2+1), \qquad \phi_{j}(0) = \lambda_{j}^{-1}, 
\end{align}
with $\lambda_{j} \in (0,1)$.
As~\cite{Carnevale.2007}, it can be easily shown that $\phi_{j}(\tau) \in [\lambda_{j},\lambda_{j}^{-1}]$ for all $\tau \in [0,T_j]$.

{Similar to \cite{Abdelrahim.2016}}, define the hybrid Lyapunov function $U (\xi) := V(x) + \max \{ \gamma \phi_{j}(\tau)W^2(\kappa,e),0 \}$,  
for all $\xi \in \C \cup \D$.
It is follows from the definition of $U$,~\eqref{eq:e17} and~\eqref{eq:e19} that $\underline\alpha_x(|x|) \leq U(\xi) \leq \overline\alpha_x(|x|) + \gamma \lambda_{\min}^{-1} \overline\alpha_e^2(|e|)$, where  $\lambda_{\min}:= \min\limits_j \lambda_j$.

Let $\xi \in \C$ and consider two different cases: i) If $\phi_{j}(\tau) \geq 0$, 
it follows from \eqref{eq:e18},~\eqref{eq:e21} and~\eqref{eq:e25} (after some calculation) that $U^{\circ} (\xi;F(\xi)) \leq  - \eta V(x) - \eta W^2(\kappa,e) \leq \! - \ol\eta  U(\xi),$
where $\ol\eta := \eta \min\{1,\lambda_{\min}/\gamma\}$.
ii) If $\phi_{j}(\tau)<0$, according to~\eqref{eq:e18}, it is derived that $U^{\circ} (\xi;F(\xi)) \leq - \eta V(x) - (\eta - \gamma^2) W^2(\kappa,e) - H^2(x).$  
Since $\phi_{j}(\tau) \in [\lambda_{j},\lambda_{j}^{-1}]$ for all $\tau \in [0,T_j]$, it is concluded from $\phi_{j}(\tau)<0$ that $\tau > T_j$.
From this fact and based on the definition of $\C$, it is inferred that $|e| \leq d$.
Using this and also~\eqref{eq:e19} in the the last inequality we get $U^{\circ} (\xi;F(\xi)) \leq - \eta U(\xi) + (\gamma^2 - \eta)\overline \alpha_e^2(d)$.
By the results of cases i) and ii)  the inequality
\begin{equation}\label{eq:Udott}
U^{\circ} (\xi;F(\xi)) \leq  - \eta U(\xi) + \hat d, 
\end{equation}
holds for all $\xi \in \C$, with $\hat d = (\gamma^2 - \eta)\overline \alpha_e^2(d)$. 

Moreover, for any $\xi \in \mathcal{D}$ we have $U (\xi^+) = V(x^+) + \gamma \phi_{j}(\tau^+) [W(\kappa^+,e^+)]^2$.
It follows from \eqref{ncs-hybrid},~\eqref{eq:e20} and the fact $\phi(\tau^+) = \lambda^{-1}_{j+1}$ that $U (\xi^+) \leq V(x) + \gamma \lambda^{-1}_{j+1} [\sigma (W(\kappa,e)) ]^2$.
From the generation of $\{\lambda_{j}\}_{j \in \Zp}$, we have that $ U (\xi^+)  \leq V(x) + ({\lambda_{j+1}}/{\lambda_{j}}) \gamma \lambda_{j} [W (\kappa,e)]^2. $
Since $\phi_{j}(\tau) > \lambda_{j}$, it is concluded that $ U (\xi^+)  \leq V(x) + ({\lambda_{j+1}}/{\lambda_{j}}) \gamma \phi_{j}(\tau) [W (\kappa,e)]^2. $ If $\lambda_{j+1}/ \lambda_{j} < 1$ it is trivially concluded that $U(\xi^+) \leq U(\xi)$.
Now assume that $\lambda_{j+1}/ \lambda_{j} \geq 1$.
Then, from the last inequality the following holds
$ U (\xi^+) \leq {\lambda_{j+1}}/{\lambda_{j}} U(\xi)$.
Partition $E := \bigcup\limits_{k = 0}^{J} {\left( {\left[ {{t_k},{t_{k + 1}}} \right],k} \right)}$ with $t_0 = 0$, $t \in [t_{J},t_{J+1}]$ and $U_0 := U(0,0)$.
Concatenating flows and jumps yields 
\begin{align} \label{eq:e86}
& U (t,j) \leq  \bar U_0 e^{- \eta (t/2 + \epsilon j/2)} + \tilde d
\end{align} 
for all $(t,j) \in E$, where $\epsilon$ is the minimum inter-jump time and $\bar U_0 = \frac{\lambda_{\max}}{\lambda_{\min}}U(0,0)$, $\tilde d = \frac{ \lambda_{\max}\hat d}{\lambda_{\min} \eta (1-e^{-\eta \epsilon})}$, $\lambda_\mathrm{max}:=\max_j\lambda_j$.

By the lower and upper bounds on $U$, and~\eqref{eq:e86}, we conclude~\eqref{eq:GApS} with $\beta(s,t,j) = \underline \alpha_x^{-1}(2 \overline \alpha(s) e^{-\eta(t/2+\epsilon j/2)})$, and $\delta = 2 \tilde d$, where $\overline\alpha(\cdot) := \lambda_{\max}/\lambda_{\min}[\overline\alpha_x(\cdot) + \gamma \lambda_{\min}^{-1} \overline\alpha_e^2(\cdot)]$.
\hfill $\Box$
\end{document}